\documentclass[aps,prb,twocolumn,floatfix,superscriptaddress,longbibliography,showpacs,nofootinbib]{revtex4-2}

\usepackage[utf8]{inputenc}
\usepackage{graphicx}
\usepackage{dcolumn}
\usepackage{bm}
\usepackage{amsfonts}
\usepackage{amsmath}
\usepackage{amssymb}
\usepackage{braket}
\usepackage{color,soul}
\usepackage{wasysym}
\usepackage{mathrsfs}
\usepackage{times}
\usepackage[dvipsnames,svgnames,table]{xcolor}
\usepackage{soul}
\usepackage{hyperref}
\usepackage{fancyhdr}
\usepackage[english]{babel}
\usepackage{longtable}
\usepackage{multirow}
\usepackage{float}
\usepackage{comment}
\usepackage{units}
\usepackage{makecell}
\usepackage{enumitem}
\usepackage{circledsteps}

\hypersetup{
    unicode=true,			
    pdftoolbar=true,			
    pdfmenubar=true,			
    pdffitwindow=false,			
    pdfstartview={FitH},		
    pdfauthor={FDR},			
    colorlinks=true,			
    linkcolor=NavyBlue,			
    citecolor=Maroon,			
    filecolor=NavyBlue,			
    urlcolor=NavyBlue       		
}

\begin{document}

\title{Hysteresis and effective reciprocity breaking due to current-induced forces}

\author{Erika L. Mehring}
\affiliation{Instituto de Física Enrique Gaviola (CONICET) and FaMAF, Universidad Nacional de Córdoba, Argentina}

\author{Raúl A. Bustos-Marún}
\affiliation{Instituto de Física Enrique Gaviola (CONICET) and FaMAF, Universidad Nacional de Córdoba, Argentina}
\affiliation{Facultad de Ciencias Químicas, Universidad Nacional de Córdoba, Argentina}

\author{Hernán L. Calvo}
\affiliation{Instituto de Física Enrique Gaviola (CONICET) and FaMAF, Universidad Nacional de Córdoba, Argentina}

\begin{abstract}
Directed transport is a key concept for many ongoing applications including nanoscale heat management, current rectification, source protection, and energy harvesting. Within the context of quantum transport, we here explore the use of nonlinear effects introduced by current-induced forces (CIFs) as a practical way to effectively break charge and heat transport reciprocities. In particular, we consider a simple model consisting of a mobile quantum dot (QD) coupled to two leads, where the charge (or heat) current develops an asymmetric behavior under inversion of voltage (or temperature) bias, thereby turning the system into a quantum diode (or quantum thermal diode). Furthermore, we find multiple stable positions for the QD and we show how the extraction of useful work is possible by modulating the nonequilibrium sources along well-established hysteresis loops. Finally, we explore a particular case where the nonlinearity of the CIFs can be exploited to pump heat or charge, even for systems that preserve inversion symmetry. This counterintuitive result is attributed to a spontaneous breaking of the inversion symmetry due to the intrinsic system's dynamics.
\end{abstract}

\maketitle

\section{Introduction}
\label{sec:intro}

The breaking of transport reciprocity indicates that forward and backward flows differ when inverting the nonequilibrium sources (bias voltages or temperature gradients in the context of quantum transport). This phenomenon offers a way of controlling signal transport, provides isolation of sources, and is key for energy harvesting. The paradigmatic example occurs in the heterojunction of p- and n-type semiconductors (electrical diodes), and the extension of this concept to molecular diodes has been studied from the mid-70s to the present~\cite{aviram1974,diezperez2009,gupta2023}.
Nowadays, there is great interest in studying the reciprocity breaking of heat currents (a phenomenon also known as thermal rectification or thermal-diode effect), due to its importance in nanoscale heat management~\cite{chang2006,martinezperez2015,fiorino2018,senior2020}, which is particularly relevant for many quantum technologies~\cite{joulain2017,kargi2019,iorio2021,arrachea2023}.

Independently of the type of system (classical or quantum), typically two main ingredients are considered to break reciprocity: magnetism (either real or synthetic)~\cite{villegas2003,casati2007,fang2017}, and nonlinearities~\cite{terraneo2002,zeng2008,bender2013}. In the context of quantum transport, nonlinearity arises naturally due to the combined effect of confinement and electron-electron interactions, while the use of magnetic fields has been always a common ingredient within the field. In this scenario, charge and heat rectification has been widely studied in different systems, including spin chains~\cite{arrachea2009,balachandran2018}, quantum dots within the Coulomb blockade regime~\cite{ono2002,thierschmann2015,sanchez2017,aligia2020,tesser2022}, and superconductor/insulator heterojuntions~\cite{khomchenko2022,giazotto2020}. Some nanoelectromechanical systems show the appearance of a rich self-induced nonlinear dynamics, which has also been studied for charge and heat rectification. This is the case with rotary current rectifiers~\cite{waechtler2021} and parametric-instability-based electron shuttles~\cite{ahn2006,penia2013,qin2021}. Besides rectification, nonlinearities may induce other interesting phenomena such as hysteresis, which has been proposed as the basic principle for the design of novel devices~\cite{joachim2005,nilsson2006,kalita2013,negre2008,kurnosov2022}.

Current-induced forces (CIFs), sometimes also dubbed electron wind forces, emerge from the interaction between the flow of electrons (which constitutes a quantum component) and certain mechanical degrees of freedom (typically described as classical variables) that represent the nuclear motion. These forces have attracted increasing attention in many areas of condensed matter physics, due to its role in nanowires~\cite{cunningham2014,mccooey2020}, nanojuctions~\cite{zimbovskaya2020}, molecular Brownian motors~\cite{ribetto2022}, electromigration~\cite{chatterjee2018}, cooling/heating of vibrational modes~\cite{naik2006,lu2015}, and the development of several current-driven nanodevices~\cite{dundas2009,bustos2013,ludovico2016,fernandez2017,lin2019,kheradsoud2019,bustos2019,preston2023}.

Here we study the role of CIFs as a source of nonlinearity for the rectification of heat and charge currents, and their contribution to the appearance of hysteretic behaviors in molecular or nanoelectromechanical devices.
In particular, we focus on generic systems like the one depicted in Fig.~\ref{fig:scheme}, where a mobile local system (red) representing a quantum dot, molecule, or atom, is coupled to two leads. The local system is subjected to two types of forces: an equilibrium (zero bias) force, represented by the springs; and a nonequilibrium force generated by the flow of electrons through it. The nonlinearity responsible for the phenomena studied here arises from the dependence of the electronic Hamiltonian with the positions of the nuclei, which in turn depend on the CIF.

The manuscript is organized as follows. In Sec.~\ref{sec:model} we provide the general theoretical framework. In Sec.~\ref{sec:results} we show and discuss the results of the work, including: the adiabatic dynamics of the composite system (Sec.~\ref{subsec:dynamics}), the effective reciprocity breaking of charge and heat currents (Sec.~\ref{subsec:reciprocity}), the appearance of hysteresis loops and its application for work extraction (Sec.~\ref{subsec:hysteresis}), the dynamical breaking of inversion symmetry, a phenomenon that enables a form of pumping in systems with inversion symmetry (Sec.~\ref{subsec:dynamical breaking}), and the role of the force fluctuations in the nonlinear effects obtained (Sec.~\ref{sec:fluctuation}). Finally, in Sec.~\ref{sec:conclusions}, we summarize and discuss the main results.

\section{Model and Method}
\label{sec:model}

\begin{figure}[ht]
\includegraphics[width=\columnwidth]{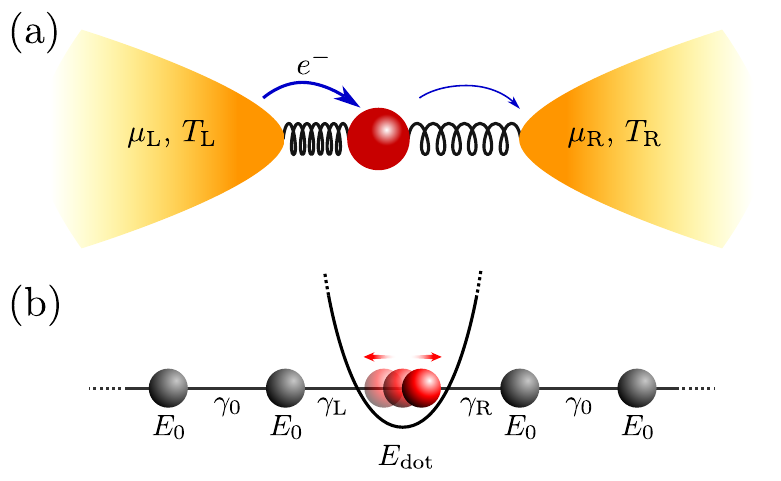}
\caption{(a) Schematics of the transport setup: a local system (red) representing a quantum dot, molecule, or atom, is coupled to two leads. The local system is allowed to move and it is subjected to a classical potential represented by the springs. The arrows represent the electron flux, which induces an additional contribution to the force exerted on the local system. (b) Hamiltonian model used to represent the system: a single site coupled, through $\gamma_\text{L}$ and $\gamma_\text{R}$ hopping amplitudes, to two semi-infinite chains with lattice parameter $a$ playing the role of the leads. The parabola accounts for the harmonic potential to which the local system is subjected.}
\label{fig:scheme}
\end{figure}

Throughout this work, we adopt a transport setup where the nanostructure of interest, which can be a molecule or a quantum dot represented by a single resonance, is bonded to two macroscopic electrodes. To this end, we consider the one-dimensional system shown in Fig.~\ref{fig:scheme}, which consists of an oscillating quantum dot coupled to left (L) and right (R) leads. The electrodes are assumed to be in equilibrium, with electrochemical potentials $\mu_\alpha = \mu_0+\delta\mu_\alpha$ and temperatures $T_\alpha = T_0 + \delta T_\alpha$, where the index $\alpha = \{\text{L},\text{R}\}$ labels the leads. Here, $\mu_0$ denotes the Fermi energy at equilibrium, $T_0$ is a temperature of reference, and for the two-lead configuration of the figure we take either $\delta\mu_\text{L} = -\delta\mu_\text{R} = eV/2$ or $\delta T_\text{L} = -\delta T_\text{R} = \delta T/2$, where $e$ is the electron charge unit, and $V$ ($\delta T$) is the applied voltage (temperature) bias. The electronic system is described by the following nearest-neighbor tight-binding Hamiltonian:
\begin{equation}
\hat{H}_\text{el} = \sum_i E_i \hat{c}_{i}^\dag \hat{c}_i - \sum_{\braket{i,j}} \gamma_{i,j} \hat{c}_{i}^\dag \hat{c}_j,
\end{equation}
where $\hat{c}_{i}^\dag$ ($\hat{c}_i$) creates (annihilates) an electron at site $i$, with onsite energy $E_i$, $\gamma_{i,j}$ represents the hopping amplitude between sites $i$ and $j$, and $\braket{i,j}$ means that the sum is carried out over nearest neighbor sites. This Hamiltonian describes the one-dimensional chain shown in Fig.~\ref{fig:scheme}(b), where the central site represents the oscillating molecule or quantum dot (QD from now on) and the leads are modeled by two semi-infinite chains. The QD is allowed to move in the transport direction, and its displacement from the equilibrium position is described through the mechanical coordinate $\hat{x}$. We will work under the nonequilibrium Born-Oppenheimer approximation~\cite{bode2011} (or Ehrenfest approximation~\cite{dundas2009,cunningham2014}) as we treat $x$ as a classical variable. Under this assumption, $x$ represents the expectation value of the position operator, neglecting its quantum fluctuations $\xi_x$, i.e., $x=\braket{\hat x}+\xi_x \approx \braket{\hat x}$. We define $x_0$ as the natural equilibrium position of the QD, i.e., the dot's equilibrium position in the absence of an applied bias voltage or temperature gradient.
By considering $a_0$ the distance between the QD equilibrium position $x_0 \equiv 0$ and the leads, we assume for the tunnel couplings the following functions:
\begin{equation}
\gamma_\text{L} = \gamma (1+\lambda) e^{-bx/a_0}, \, \gamma_\text{R} = \gamma (1-\lambda) e^{+bx/a_0},
\label{eq:hop}
\end{equation}
where $b$ is a constant which describes the electromechanical coupling strength. Moreover, $\gamma_\text{L}$ and $\gamma_\text{R}$ depend on the asymmetry factor $-1<\lambda<1$. Within the leads we set the onsite energies as $E_i = E_0$ and the hopping amplitudes as $\gamma_0$, while for the QD we use $E_\text{dot}$.

The dynamics of the QD obeys a Langevin equation:
\begin{equation}
M \ddot{x} + \partial_x U(x,t) + \xi = F_\text{el}.
\label{eq:langevin}
\end{equation}
Here, $M$ is the mass associated with the mechanical system, and $F_\text{el}$ represents the electronic force to which the QD is subjected. This force is calculated as the expectation value of the electronic Hamiltonian’s variation with respect to $x$, i.e., $F_\text{el} = \braket{-\partial_x \hat{H}_\text{el}}$, and it arises from the interaction between the electrons and the mechanical device. The term $\partial_x U(x,t)$ accounts for any mechanical force not included in $F_\text{el}$, and its time dependence is a liberty in notation, indicating possible (classical) nonconservative effects, such as mechanical friction.
Additionally, $\xi$ describes stochastic fluctuations of the forces. This quantity presents at least two contributions: the electronic stochastic force and a mechanical fluctuation which can be attributed, e.g., to the coupling between the QD and a phonon bath. In this work, we will consider a large mechanical fluctuation compared with its electronic counterpart, and we will relate it, through fluctuation-dissipation theorem, to a mechanical friction force. For clarity in the discussions, we first explore the \textit{average} effect of the current-induced forces (CIFs) by neglecting $\xi$. Then, in Sec.~\ref{sec:fluctuation}, we discuss possible effects introduced by the fluctuation term.

In the nonequilibrium Green's function (NEGF) formalism, we can compute the electronic force by using:
\begin{equation}
F_\text{el} = -i \hbar \, \text{Tr}\left[\bm{\Lambda}\bm{\mathcal{G}}^< (t,t)\right],
\label{eq:force_el}
\end{equation}
where the matrix $\bm{\Lambda} = -\partial_x \bm{H}_\text{el}$ is the force operator written in a given basis and the matrix $\bm{\mathcal{G}}^< (t,t)$ is the lesser Green's function accounting for the nonequilibrium, single-particle density matrix $\bm{\rho}(t)$ of the electrons, i.e., $\bm{\rho}(t)=-i\hbar \bm{\mathcal{G}}^< (t,t)$. This force can be linearly decomposed in terms of equilibrium and nonequilibrium contributions
\begin{equation}
F_\text{el} = F^\text{eq}+F^\text{ne}.
\end{equation}
In our model, we suppose that $F^\text{eq}$ is already included in the conservative part of $\partial_x U$, in the form of a harmonic force. The nonequilibrium term, $F^\text{ne}$, is the so-called CIF.
\footnote{The electronic forces are dubbed by some authors as current-induced forces, where the general expressions can be applied to both equilibrium and nonequilibrium situations~\cite{bode2011,bustos2013,ribetto2021}. Here we preserve this term only for conditions (out of equilibrium) such that particle (or heat) current is possible.}
The contributions $F^\text{eq}$ and $F^\text{ne}$ are also dubbed ``direct force'' and ``electron wind force'', respectively, in other
contexts~\cite{vogel2017}. With these considerations, we rewrite Eq.~(\ref{eq:langevin}) as follows:
\begin{equation}
M \ddot{x} + kx + \nu \dot{x} = F^\text{ne}(x),
\label{eq:motion}
\end{equation}
where $k$ is the force constant associated with the harmonic potential shown in Fig.~\ref{fig:scheme} and $\nu$ is the mechanical friction coefficient arising from the dissipative contribution of $\partial_x U$. This coefficient can be related to the (dimensionless) damping ratio $\zeta$ through $\nu = 2\zeta \sqrt{M k}$, and we will work within the underdamped regime $\zeta<1$. In the following, we will calculate $F^\text{ne}$ through Eq.~(\ref{eq:force_el}) and solve Eq.~(\ref{eq:motion}) in the adiabatic limit, where the mechanical coordinate moves much more slowly than the electrons flowing through the QD. Note that since the CIF depends parametrically on $x$, the integration of Eq.~(\ref{eq:langevin}) is performed in a self-consistent way.

Regarding the scales for the quantities involved in Eq.~(\ref{eq:motion}), we notice that the one-dimensional chains modeling the leads yield a natural length given by the lattice parameter $a$, so the typical unit for the mechanical coordinate can be chosen through this quantity.~\footnote{For the homogeneous chain and setting the Fermi energy as $\mu_0 = E_0$, the lattice parameter results to be $a = \hbar v_\text{F}/2\gamma_0$. In a more general model for the contacts, it is possible to relate the coordinate unit with the Fermi velocity as $x^* \simeq v_\text{F} t^*$ where $t^*$ is the electron's time unit.} For the CIF one can observe that, from its definition as the derivative of the electronic energy with respect to $x$, that it scales as $\gamma_0/a$, while the typical time unit for the electron dynamics is given by $t^* = \hbar/\gamma_0$. From these estimations, in Appendix~\ref{app:unit} we derive the typical scales for the remaining quantities.

In the adiabatic limit, we can expand the exact Green's function $\bm{\mathcal{G}}^<(t,t)$ in terms of the frozen Green's function $\bm{G}^<(t,t)$. Here, the term ``frozen'' refers to the description of the same quantity but calculated through the time-independent Hamiltonian that results from setting the parameters at time $t$~\cite{bode2011,deghi2021}.
In other words, $\bm{G}^<(t,t)$ is the Born-Oppenheimer approximation of $\bm{\mathcal{G}}^<(t,t)$ and, as a result, it depends only \textit{parametrically} on time.
At zeroth order of the adiabatic expansion and within the NEGF (Keldysh) formalism, the lesser Green's function $\bm{\mathcal{G}}^<(t,t)$ yields
\begin{equation}
\bm{\mathcal{G}}^{<}(t,t) \approx \bm{G}^{<}(t,t) = \int \frac{\text{d}\epsilon}{2\pi\hbar} \bm{G}^r(\epsilon) \bm{\Sigma}^{<}(\epsilon) \bm{G}^a(\epsilon),
\label{eq:g^<}
\end{equation}
where $\bm{G}^r(\epsilon)$ and $\bm{G}^a(\epsilon)=[\bm{G}^r(\epsilon)]^\dag$ are the Wigner transforms of the frozen retarded and advanced Green's function matrices, respectively, and $\bm{\Sigma}^{<}(\epsilon)$ is the energy domain lesser self-energy. The latter can be written in terms of the Fermi distribution function $f_\alpha(\epsilon)=[1 + \exp(\epsilon-\mu_\alpha)/k_\text{B} T_\alpha]^{-1}$ of the reservoir $\alpha$ (at temperature $T_\alpha$ and chemical potential $\mu_\alpha$) as
\begin{equation}
\bm{\Sigma}^{<}(\epsilon) = 2i \sum_\alpha f_\alpha(\epsilon) \bm{\Gamma}_\alpha(\epsilon),
\end{equation}
where $\bm{\Gamma}_\alpha = (\bm{\Sigma}^{a}_\alpha - \bm{\Sigma}^{r}_\alpha)/2i$ represents the escape rate of the wave function through the $\alpha$-lead, and $\bm{\Sigma}^{r}_\alpha$ ($\bm{\Sigma}^{a}_\alpha$) is the $\alpha$-lead contribution to the total retarded (advanced) self-energy. Note that this $\bm{\Gamma}_\alpha$ and the one defined in other conventions~\cite{gaury2014} differ in a factor of 2. Replacing the above expressions in Eq.~(\ref{eq:force_el}), we obtain:
\begin{equation}
F_\text{el} = \int \frac{\text{d}\epsilon}{\pi} \sum_\alpha f_\alpha \text{Tr}[\bm{\Lambda}\bm{G}^r \bm{\Gamma}_\alpha \bm{G}^a],
\label{eq:force_gral}
\end{equation}
where we omitted the energy argument $\epsilon$ to keep the notation simple.
In the linear response regime, the Fermi functions $f_{\alpha}$ are expanded up to first order in terms of $\delta \mu_\alpha$ and $\delta T_\alpha$, according to
 \begin{equation}
f_{\alpha} \approx	f_0 - \frac{\partial f_0}{\partial\epsilon} \, \delta\mu_\alpha - (\epsilon-\mu_0) \, \frac{\partial f_0}{\partial\epsilon} \, \frac{\delta T_\alpha}{T_0},
\end{equation}
where $f_0$ is the Fermi function with $\mu_\alpha=\mu_0$ and $T_\alpha=T_0$.
Given that the energy derivative of the Fermi function is different from zero only in the  vicinity of $\mu_0$ (within the order $k_\text{B}T_0$), an expansion of the force kernels
$\mathcal{F}_\alpha = \text{Tr}[\bm{\Lambda} \bm{G}^r \bm{\Gamma}_\alpha \bm{G}^a]$ in terms of $\epsilon$ naturally holds. When these kernels vary smoothly within this range, the energy expansion is sufficient up to first-order, i.e.,
 \begin{equation}
\mathcal{F}_\alpha(\epsilon) \approx \mathcal{F}_\alpha(\mu_0) + \left.\frac{\partial\mathcal{F}_\alpha}{\partial\epsilon}\right|_{\mu_0}(\epsilon-\mu_0).
\end{equation}
Using this, the CIF can be decomposed in terms of the voltage and temperature bias contributions~\cite{deghi2021}: $F^\text{ne}=F_{\delta\mu}^\text{ne}+F_{\delta T}^\text{ne}$, where
\begin{align}
F_{\delta\mu}^\text{ne} &=  \frac{1}{\pi} \sum_\alpha \delta\mu_\alpha \mathcal{F}_\alpha(\mu_0),  \label{eq:cif_V}\\
F_{\delta T}^\text{ne} &= \frac{\pi}{3} \left(k_\text{B}T_0\right)^2 \sum_\alpha \frac{\delta T_\alpha}{T_0} \left. \frac{\partial \mathcal{F}_\alpha}{\partial \epsilon}\right|_{\mu_0} .
\label{eq:cif_T}
\end{align}
By working on the site basis, the retarded Green's function $\bm{G}^r$ can be obtained in the usual way:
\begin{equation}
\bm{G}^r = \left[(\epsilon+i0^+)-\bm{H}_\text{eff}\right]^{-1},
\label{eq:Green}
\end{equation}
where the presence of the leads in the electronic Hamiltonian is accounted for by semi-infinite homogeneous chains. Therefore, an effective Hamiltonian $\bm{H}_\text{eff}$ can be obtained by considering a three-site system where the central site represents the QD and the remaining ones include the retarded self-energy corrections $\Sigma^r_\alpha$ due to the presence of the leads. Hence, the electronic Hamiltonian can be written as the following matrix:
\begin{equation}
\bm{H}_\text{eff} = \begin{pmatrix}
E_0+\Sigma^r_\text{L} & -\gamma_\text{L} & 0 \\
-\gamma_\text{L} & E_\text{dot} & -\gamma_\text{R} \\
0 & -\gamma_\text{R} & E_0+\Sigma^r_\text{R}
\end{pmatrix}.
\label{eq:Heff}
\end{equation}
For simplicity, we consider the semi-infinite chains on each side of the QD to be the same, such that $\Sigma^r_\text{L} = \Sigma^r_\text{R} \equiv \Sigma^r$ and we calculate it through:
\begin{equation}
\Sigma^r = \frac{\gamma_0^2}{\epsilon + i0^{+} - \Sigma^r} \rightarrow \Sigma^r = \Delta(\epsilon) - i\Gamma(\epsilon),
\end{equation}
where the real and imaginary components can be obtained from Refs.~\cite{deghi2021} and \cite{pastawski2001}.

The dependence of the force on $x$ is generally highly nonlinear. This becomes evident in the wideband limit expression found for the case of $F^\text{ne}_{\delta \mu}(x)$, as detailed in Appendix~\ref{app:formula}. There we find, for the particular case of symmetric bias:
\begin{equation}
F^\text{ne}_{\delta \mu}(x) = -\frac{eV}{\pi}\left(\frac{b}{a_0}\right)\frac{(E_\text{dot}\gamma_0)(\gamma_\text{L}^2+\gamma_\text{R}^2)}
{(E_\text{dot}\gamma_0)^2+(\gamma_\text{L}^2+\gamma_\text{R}^2)^2},
\end{equation}
where $\gamma_\text{L}^2+\gamma_\text{R}^2 = 2\gamma^2 \left[(1+\lambda^2) \cosh y - 2 \lambda \sinh y\right]$, and $y=2bx/a_0$. In addition to the linear dependence on the bias, we can see that the CIF scales with the factor $b/a_0$ and it is also proportional to the QD onsite energy. In fact, it is possible to change the sign of $F^\text{ne}$ through this energy (relative to $\mu_0$), such that for positive bias and $E_\text{dot}>0$ the CIF points to the left (hole-like behavior) while for $E_\text{dot}<0$ the CIF points to the right (electron like behavior). 

On the other hand, the analytical expression of $F^\text{ne}_{\delta T}$ is related to that of $F^\text{ne}_{\delta \mu}$, since it involves the energy derivative of the force kernel $\mathcal{F}_\alpha$, thus yielding a more intricate expression. In any case, one can numerically verify that, in general, $F^\text{ne}_{\delta T}$ displays a similar dependence on $x$ as $F^\text{ne}_{\delta \mu}$, as we illustrate in Fig.~\ref{fig:fx_dT}.

The nonlinearity of the CIFs is the origin of nontrivial effects, such as reciprocity-breaking and hysteresis, that we will address in the next sections.

\section{Results}\label{sec:results}

In the following subsections we discuss different phenomena introduced by the CIF and the nonlinear behavior it generates. Our focus, while not exclusively, centers on examples and discussions involving CIFs caused by a bias voltage, $F^\text{ne}_{\delta \mu}$, simply because these forces are in general larger and easier to tune than those arising from a temperature gradient, $F^\text{ne}_{\delta T}$. However, the nonreciprocal and hysteretic behavior obtained, together with their potential applications, can also be extended to the case where the nonequilibrium source is a temperature gradient.

\subsection{Adiabatic dynamics of the composite system}
\label{subsec:dynamics}

\begin{figure}[ht]
\includegraphics[width=\columnwidth]{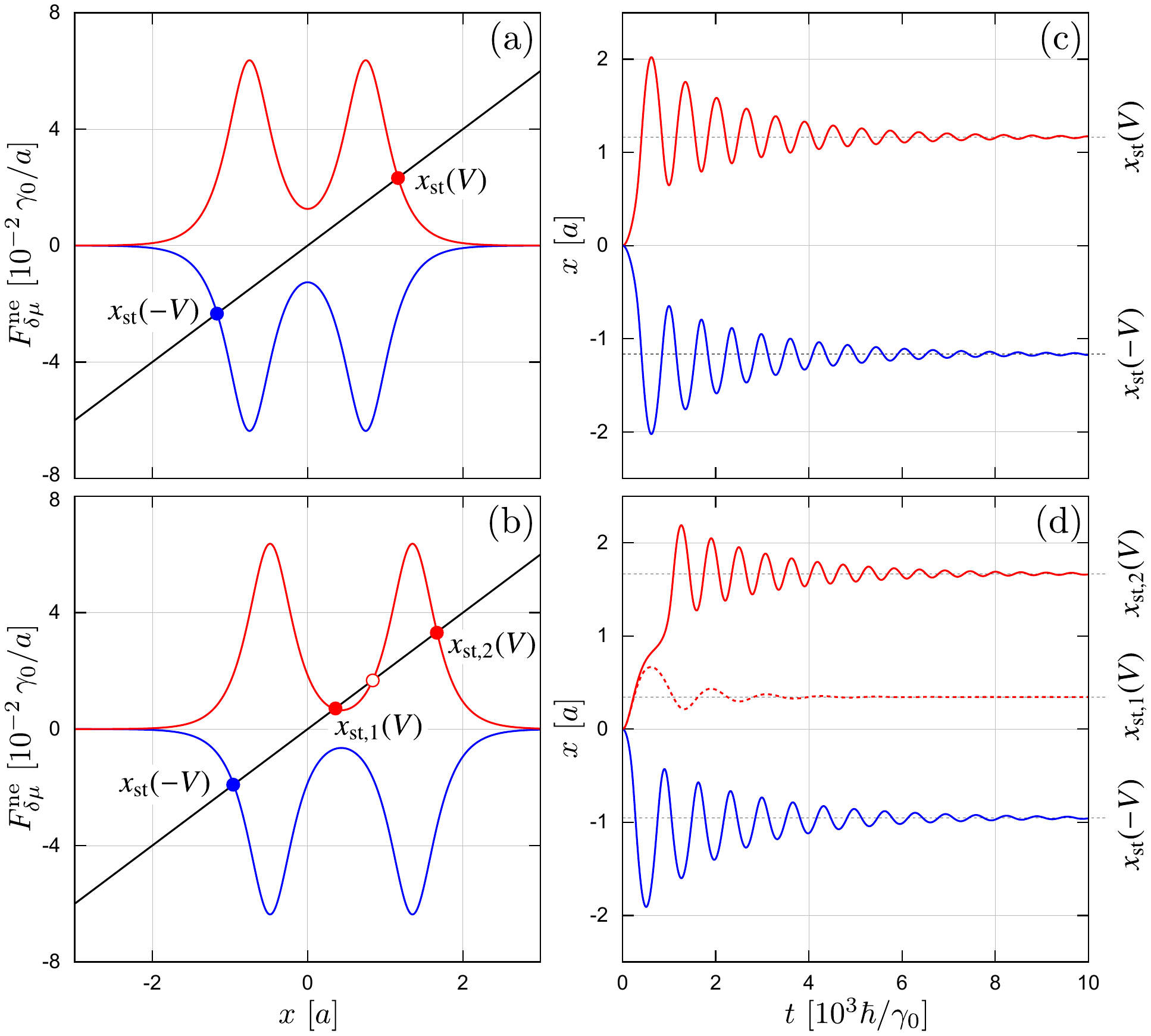}
\caption{Current induced force and adiabatic dynamics: (a,b) CIF as a function of the QD position for $eV=0.2\,\gamma_0$ (red) and $eV=-0.2\,\gamma_0$ (blue). In (a) we use an asymmetry factor $\lambda = 0$ while in (b) $\lambda = 0.7$. The straight line in black depicts the (minus) harmonic force, such that crossings with the CIF correspond to stable (filled circles) and unstable (empty circles) solutions. (c,d) Dynamics of the QD position related with $\lambda = 0$ and $\lambda = 0.7$, respectively, for an initial condition $x(0) = \dot{x}(0) = 0$. Red curves correspond to positive bias, while blue curves are for negative bias. The used parameters are: $E_\text{dot}=-0.2\,\gamma_0$, $\gamma=0.1\,\gamma_0$, $k=0.02\,\gamma_0/a^2$, $b=8$, $a_0=4a$, and for panels (c) and (d) we use $M=1000\,\hbar^2/\gamma_0 a^2$, $\zeta = 0.1$ (solid lines), and $\zeta = 0.2$ for dashed line in (d).}
\label{fig:force_dyn}
\end{figure}

As an initial illustration of the previous discussions, in this section we evaluate the CIF and the corresponding dynamics of the oscillating QD. In Fig.~\ref{fig:force_dyn}(a) and (b) we show the CIF as a function of the QD position for two values of the asymmetry factor, i.e., $\lambda=0$ (a) and $\lambda = 0.7$ (b). In each panel, we show $F^\text{ne}$ for $eV=0.2\,\gamma_0$ in red and $eV=-0.2\,\gamma_0$ in blue, alongside the (minus) harmonic force in black. The CIF displays a double peak shape, and the intersection points indicate the equilibrium positions that the QD acquires in the presence of the CIF, fulfilling the condition $kx = F^\text{ne}(x)$. This condition is satisfied in the long-time limit of Eq.~(\ref{eq:motion}), where we assume $\dot{x}(t) \rightarrow 0$ due to the presence of a friction term. The filled circles represent stable equilibrium positions $x_\text{st}$, while empty circles correspond to unstable ones. When $\lambda = 0$ we can see that the equilibrium positions associated with each one of the voltages are symmetrically located with respect to the natural equilibrium position $x_0$, i.e., $x_\text{st}(-V) = -x_\text{st}(V)$. However, this symmetry no longer holds when $\lambda \neq 0$, as can be seen in panel (b). In addition, given the shape of the CIF we can observe that it is possible to obtain multiple stable positions, i.e., $x_{\text{st},1}(V)$ and $x_{\text{st},2}(V)$ for a fixed bias voltage $V$. 

Regarding the dynamics of the oscillator, we set the harmonic force constant to $k = 0.02\,\gamma_0/a^2$ (see Appendix~\ref{app:unit}). This is done to illustrate the nonlinearities of the CIFs within the linear response regime of transport. In this case, we take a relatively small elastic constant $k \approx 0.23$ N/m, which is on the order of typical van der Waals forces between fullerene molecules~\cite{cox2007}. Although larger values of $k$ could be used, the calculation of the CIFs should be generalized to Eq.~(\ref{eq:force_gral}), to fully account for larger values of voltage or temperature biases which would be necessary to compensate for the larger values of the equilibrium force.

We show in Fig.~\ref{fig:force_dyn}(c) and (d) the time evolution of the QD position for the cases shown in (a) and (b), respectively. Again, we display in red the positive bias case, while the negative bias case is shown in blue. We can see here how the QD, starting from the natural equilibrium position $x_0$, reaches a new stable position when the bias voltage is turned on. Clearly, these equilibrium positions coincide with those obtained in panels (a) and (b). Furthermore, when the QD admits more than a single stable solution [as occurs in panel (b)], the final position will depend on the initial conditions and details of the model. In this case, we see in Fig.~\ref{fig:force_dyn}(d) that by varying the damping factor it is possible to switch $x_\text{st}$ between the two stable solutions.

\begin{figure}[ht]
\includegraphics[width=.8\columnwidth]{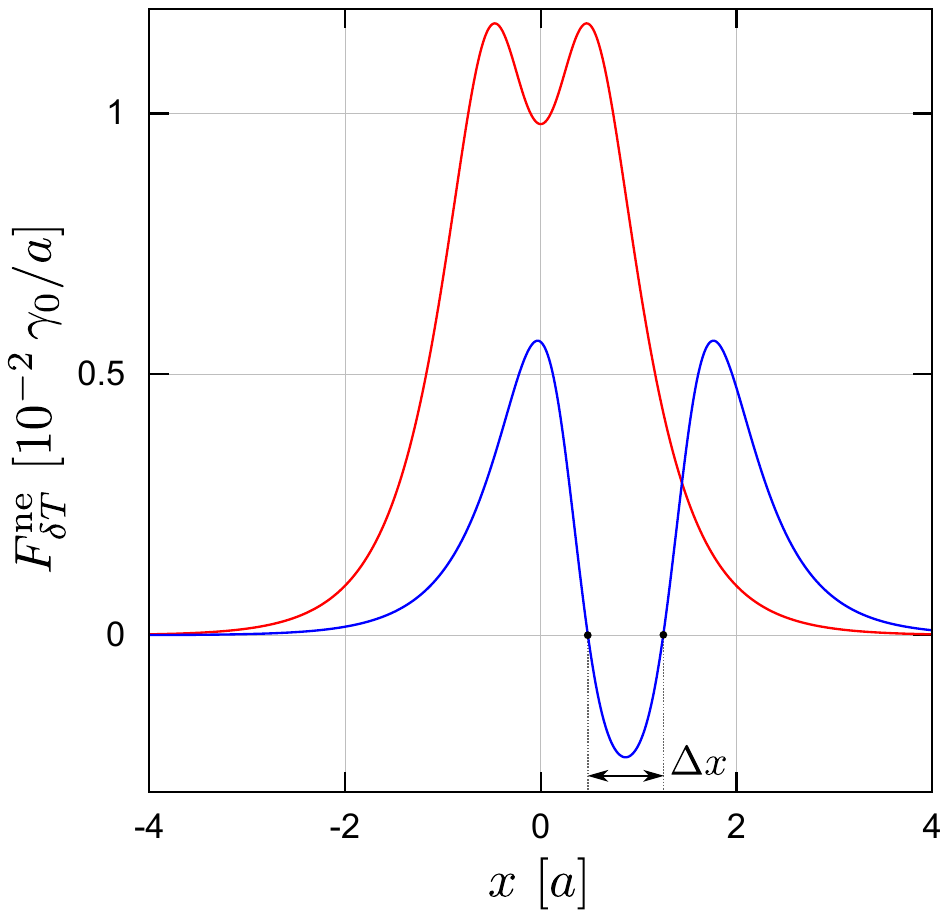}
\caption{Current induced force as a function of the QD position for $\delta T = 0.5 \, T_0$, and two coupling asymmetry factors: $\lambda = 0$ (red) and $\lambda = 0.7$ (blue). The used parameters are $E_\text{dot}=-0.3\,\gamma_0$, $\gamma=0.01\,\gamma_0$, $k=0.02\,\gamma_0/a^2$, $b=4$, $a_0=4a$, $k_\text{B}T_0 = 0.002\,\gamma_0$, and we set the Fermi energy at $\mu_0 \approx E_\text{dot}$.}
\label{fig:fx_dT}
\end{figure}

Equivalent results can be obtained for the case where the nonequilibrium source is given by a temperature gradient between the contacts. For instance, the CIF as a function of the QD position is shown in Fig.~\ref{fig:fx_dT} for two values of the asymmetry factor, $\lambda = 0$ (red) and  $\lambda = 0.7$ (blue). We observe that $F^\text{ne}_{\delta T}$ is similar to $F^\text{ne}_{\delta \mu}$ shown in Fig.~\ref{fig:force_dyn}(a) for the symmetric case, in the sense that it presents a double-peaked shape and the same sign for all values of $x$. However, for $\lambda \neq 0$ we obtain an important difference as $F^\text{ne}_{\delta T}$ changes its sign for a given range $\Delta x$ (see arrows). This can be understood through Eq.~(\ref{eq:cif_T}), since the variation of $x$ produces a shift in the resonance peak of $\mathcal{F}_\alpha$.~\footnote{Although not shown, this function typically depends on $\epsilon$ in a similar way to the transmission coefficient $\mathcal{T}$, which in the present model exhibits a peak around $E_\text{dot}$.} Therefore, if this resonance crosses the Fermi energy, a sign inversion can be expected for $F^\text{ne}_{\delta T}$. With respect to the QD dynamics, there are no significant differences worth mentioning when compared to the bias voltage case.

\subsection{Effective reciprocity breaking}
\label{subsec:reciprocity}

\textit{Charge current rectification.--} It was shown in Fig.~\ref{fig:force_dyn} that when turning on the bias voltage $V$, the oscillator reaches a steady position $x_\text{st}$ different from $x_0$. When influenced by the CIF, the oscillator moves from its natural equilibrium position $x_0$ to some other $x_\text{st}(V)$. Consequently, the charge current $I$ will also experience a corresponding change from its original value to $I(V)$. If we now invert the sign of the bias, the oscillator once more modifies its equilibrium position and the current varies to $I(-V)$. Under breaking of the coupling symmetry through $\lambda$, it is possible to see that $x_\text{st}(-V) \neq -x_\text{st}(V)$. Since the electronic current depends on the QD's final position, we can suspect some manifestation of this asymmetry in this quantity. In other words, under the bias inversion operation, the two currents are not necessarily antisymmetric, i.e., $I(-V)\neq -I(V)$, reflecting an effective reciprocity breaking in the transport due to a cooperative effect between the coupling asymmetry and the QD's motion into different steady-state positions as a function of the bias voltage. Encouraged by this result, we evaluate the rectification factor $R_I$ of the system, here defined as:
\footnote{For clarity in the figures, here we define $R_I$ and $R_J$ without the usual absolute value, see Ref.~\cite{roberts2011}.}
\begin{equation}
R_I(V) = \frac{I(V)+I(-V)}{I(V)-I(-V)}.
\end{equation}
In the linear response regime, the current is given by the transmission coefficient $\mathcal{T}_\text{RL} = \mathcal{T}_\text{LR} = \mathcal{T}$ through Landauer's formula $I \equiv (I_\text{L}-I_\text{R})/2 = (e^2/h)\mathcal{T}V$. Certainly, while these coefficients are reciprocal for a fixed position of the QD, the reciprocity breaking of the current arises from the fact that $\mathcal{T}$ ultimately depends on $x_\text{st}$. In this limit, the rectification factor reads
\begin{equation}
R_I(V) = \frac{\mathcal{T}(V)-\mathcal{T}(-V)}{\mathcal{T}(V)+\mathcal{T}(-V)}.
\label{eq:r_IV}
\end{equation}
For a given value of the applied bias, the equilibrium position of the oscillator is obtained when the sum of the forces acting on the QD becomes zero. This condition implies $F^\text{ne}(x_\text{st}) = kx_\text{st}$,  see Eq.~(\ref{eq:motion}). Once we obtain $x_\text{st}$, the transmission coefficient can be calculated from the Fisher-Lee formula~\cite{fisher1981}:
\begin{equation}
\mathcal{T}_{\alpha\beta} = \text{Tr}[2\bm{\Gamma}_\alpha \bm{G}^r 2\bm{\Gamma}_\beta \bm{G}^a],
\end{equation}
where the $x$-dependence comes from the retarded and advanced Green's functions (and not from the $\Gamma$'s), as they depend on the effective Hamiltonian given in Eq.~(\ref{eq:Green}).

\begin{figure*}[ht]
\includegraphics[width=\textwidth]{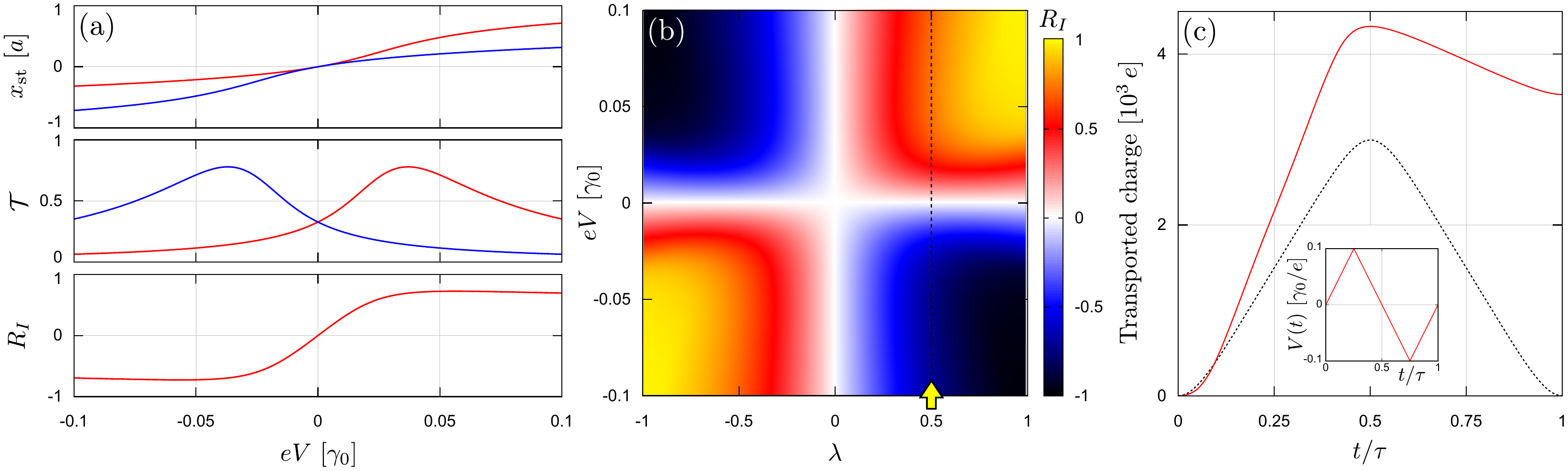}
\caption{Effective reciprocity breaking in the transport current. (a)
QD's equilibrium position $x_\text{st}$, transmission coefficient $\mathcal{T}$, and rectification factor $R_I$ as functions of the bias voltage $V$ for a fixed coupling asymmetry $\lambda = 0.5$. Going from top to bottom panels, we show $x_\text{st} (V)$, $\mathcal{T}(V)$ and $R_I(V)$ in solid red lines, while $-x_\text{st}(-V)$ and $\mathcal{T}(-V)$ are marked in solid blue lines. (b) Map of the rectification factor $R_I$ as a function of $\lambda$ and $V$, the arrow denotes the $\lambda$ value used in panel (a). (c) Transported charge through the QD as a function of time for $\lambda = 0.5$ (solid red) and $\lambda = 0$ (dashed black). The inset shows the bias voltage profile along one driving period. The used parameters are $E_\text{dot}=-0.2\,\gamma_0$, $\gamma=0.5\,\gamma_0$, $k=0.02\,\gamma_0/a^2$, $b=6$,  $a_0=4a$, and $\mu_0 = E_0 = 0$. For the dynamics in panel (c) we use $\tau = 10^6 \, \hbar/\gamma_0$, $M = 1000 \, \hbar^2/\gamma_0 a^2$, and $\zeta = 0.5$.}
\label{fig:bias1}
\end{figure*}

The rectification factor can range from $R_I=-1$ to $R_I=1$, sensing the asymmetric behavior of the transport current under bias inversion. In particular, $R_I=0$ signals the case where the current only switches its sign with the $V \rightarrow -V$ operation, but its magnitude remains unchanged. This result is expected in a symmetrically coupled system where $\lambda = 0$, such that the displacement of the mechanical oscillator obeys $x_\text{st}(V) = -x_\text{st}(-V)$. To achieve a nonzero value of $R_I$ in this regime with a single stable solution, it is necessary to break this coupling symmetry. This can be done, for example, by either setting $\lambda$ to a nonzero value or by shifting the natural QD equilibrium position $x_0$ towards one of the two contacts.

In Fig.~\ref{fig:bias1}(a) the QD position $x_\text{st}$, transmission coefficient $\mathcal{T}$, and rectification factor $R_I$ are shown as functions of the applied bias voltage $V$ with a fixed $\lambda = 0.5$. For the QD position, the comparison between $x_\text{st}(V)$ (solid red) and $-x_\text{st}(-V)$ (solid blue) reflects that $x_\text{st}$ is not an odd function of $V$. Since at zero bias the CIF is zero, this result reveals the nonlinear dependence of the CIF on the QD position. Additionally, due to the used $\lambda>0$ asymmetry, the displacement of the QD from $x_0$ is more sensitive for positive bias than negative ones. A similar behavior is observed for the charge current (through the transmission coefficient), evidencing the strong impact that the displacement of the QD has on this quantity. Finally, we plot the rectification factor $R_I$ obtained from Eq.~(\ref{eq:r_IV}). It is observed that, in the used parameter regime, this factor varies almost linearly for small applied biases and then it reaches a maximum value $R_I \approx 0.7$ for $eV \approx 0.05\,\gamma_0$. In Fig.~\ref{fig:bias1}(b) we extend this plot for different values of $V$ and $\lambda$, where some regions with $|R_I| \approx 1$ can be seen, such that the device could operate as an almost perfect quantum diode~\cite{zhang2017}.

To illustrate the breaking of the current reciprocity, we consider a slow modulation of the bias voltage, such that in a period $\tau \gg t^*$ of the modulation $V$ sweeps the range $[-0.1,0.1] \, \gamma_0$, spending the same amount of time for positive and negative values. By taking the adiabatic limit in which the time variation of $V$ is much slower than the typical response time of the QD, one can calculate the transported charge from the time-integral of the stationary charge current, i.e.,
\begin{equation}
Q(t) = \int_0^t \text{d}t' I[x_\text{st}(t')].
\label{eq:pump}
\end{equation}
This can be safely done by initially placing the QD in a stationary position for a given $V$ and taking a large $\tau$, such that the QD velocity can be neglected in a first approximation. In Fig.~\ref{fig:bias1}(c) we calculate this time-integral for the bias profile shown in the inset, such that for $t=\tau$ one obtains the net transported charge in a cycle. Here, the rectification effect on the current is clearly visible. The case where $\lambda = 0$ (dashed black) shows that, for inversion-symmetric systems, there is no net transported charge. Conversely, for $\lambda \neq 0$ (solid red) this quantity is different from zero and proportional to $\tau$.\\

\begin{figure}[ht]
\includegraphics[width=0.8\columnwidth]{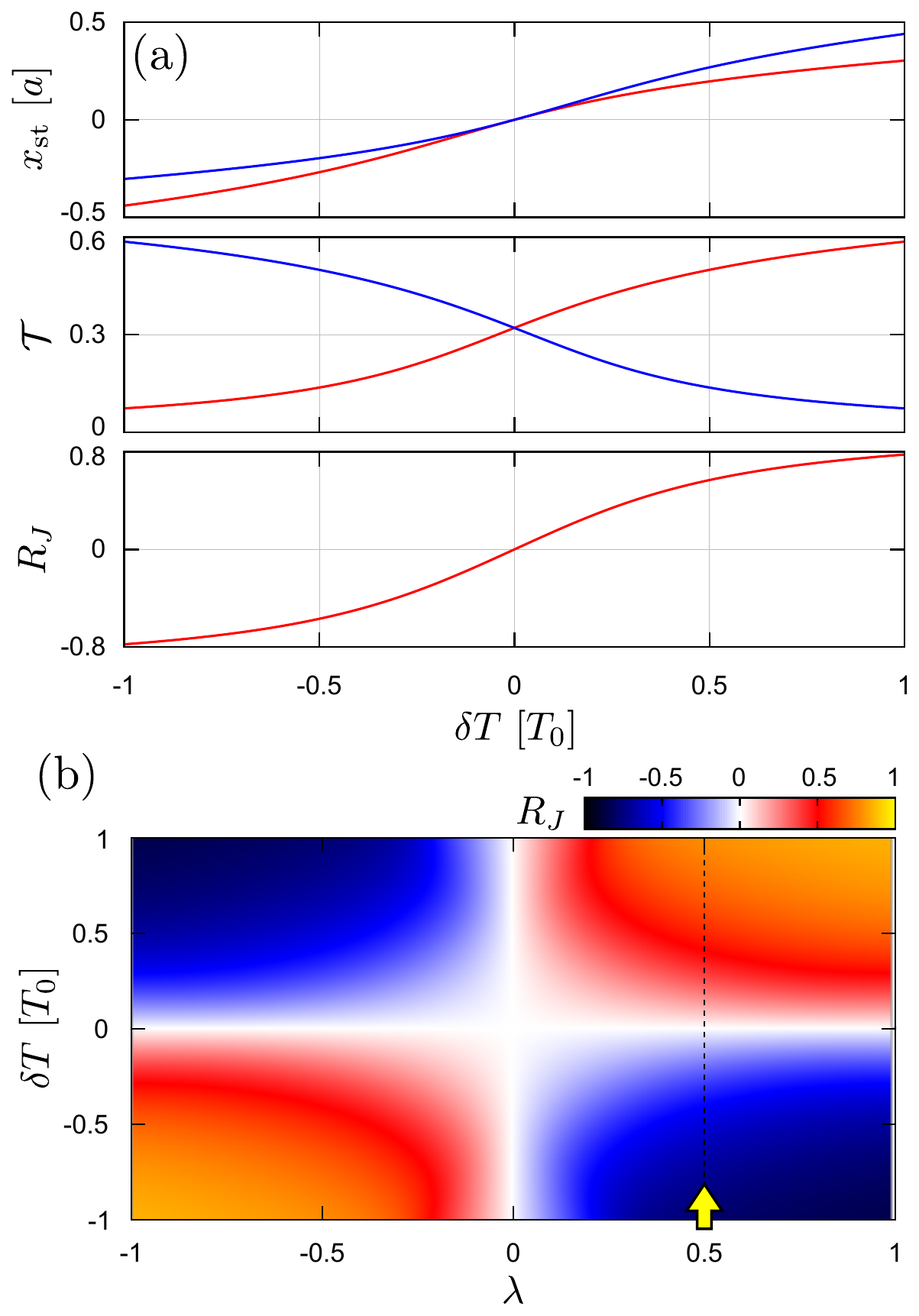}
\caption{Effective reciprocity breaking in the heat current. (a)
QD's equilibrium position $x_\text{st}$, transmission coefficient $\mathcal{T}$, and heat rectification factor $R_J$ as functions of the temperature gradient $\delta T$ for a fixed coupling asymmetry $\lambda = 0.5$. Going from top to bottom panels, we show $x_\text{st}(\delta T)$, $\mathcal{T}(\delta T)$ and $R_J(\delta T)$ in solid red lines, while $-x_\text{st}(-\delta T)$ and $\mathcal{T}(-\delta T)$ are marked in solid blue lines. (b) Map of the rectification factor $R_J$ as a function of $\lambda$ and $\delta T$, the arrow denotes the $\lambda$ value used in panel (a). The other parameters remain the same as those used in Fig.~\ref{fig:fx_dT}.} 
\label{fig:temp1}
\end{figure}

\textit{Heat current rectification.--} Since the above-described reciprocity breaking is simply related to nonequilibrium forces acting upon the mechanical degree of freedom, here we explore the possibility of using $F_{\delta T}^\text{ne}$ to make a quantum thermal diode~\cite{joulain2017,kargi2019}.
To this end, we can extend the above expressions for the rectification factor to the case where a symmetric temperature gradient $\delta T_\text{L} = -\delta T_\text{R} = \delta T/2$ is applied between the contacts. In this case, we analyze the heat current $J$ flowing through the QD system, such that the heat current rectification factor $R_J$ reads:
\begin{equation}
R_J(\delta T) = \frac{J(+\delta T)+J(-\delta T)}{J(+\delta T)-J(-\delta T)}.
\label{eq:r_JT}
\end{equation}
In the linear response regime the heat current can be computed through the transmission coefficient as
\begin{equation}
J(\delta T) = \frac{(\pi k_\text{B}T_0)^2}{3h}\mathcal{T}(\delta T)\frac{\delta T}{T_0},
\label{eq:heat_current}
\end{equation}
where $\mathcal{T}$ depends on $\delta T$ through the QD's position, i.e., $\mathcal{T}(\delta T) \equiv \mathcal{T}[x_\text{st}(\delta T)]$. Therefore, if we now calculate the rectification factor for the heat currents, we obtain Eq.~(\ref{eq:r_IV}) with the obvious change $V \rightarrow \delta T$. 

In Fig.~\ref{fig:temp1} we show the analogous of Fig.~\ref{fig:bias1} when the CIF is induced by a temperature gradient $\delta T$. It is clear that the main features discussed in the previous case at finite $V$ are also present here. Thus, the proposed model presents an effective breaking of the heat current reciprocity. Notably, even in this linear transport regime, the system can exhibit strong thermal reciprocity breaking with relatively large values of $R_J$, such that the device can also operate as a quantum thermal diode. It is important to highlight the system's tunability of its parameters, that allows both $R_J$ and $\mathcal{T}$ to reach relatively large values, as can be seen in Fig.~\ref{fig:temp1}(a). In this way, other measures of the performance of the heat rectifier will also result in favorable outcomes~\cite{khandelwal2023}.

Regarding the CIF for an applied thermal bias, c.f. Eq.~(\ref{eq:cif_T}), it is worth mentioning that for low temperatures this quantity can be significantly smaller compared to the one obtained for a bias voltage, due to its quadratic dependence on the thermal energy $k_\text{B} T_0$. However, since this quantity involves the energy derivative of the force kernel $\mathcal{F}_\alpha$, it presents a well defined two-peaked shape if the QD is weakly coupled to the leads, i.e., $\gamma \ll \gamma_0$, see Fig.~\ref{fig:fx_dT}. Therefore, a fine-tuning of the Fermi energy at one of the peaks of $\partial \mathcal{F}_\alpha / \partial \epsilon$ can increase the final value of the CIF, such that it would become comparable to that of Eq.~(\ref{eq:cif_V}).

\subsection{Hysteresis and work extraction}
\label{subsec:hysteresis}

\begin{figure*}[ht]
\includegraphics[width=\textwidth]{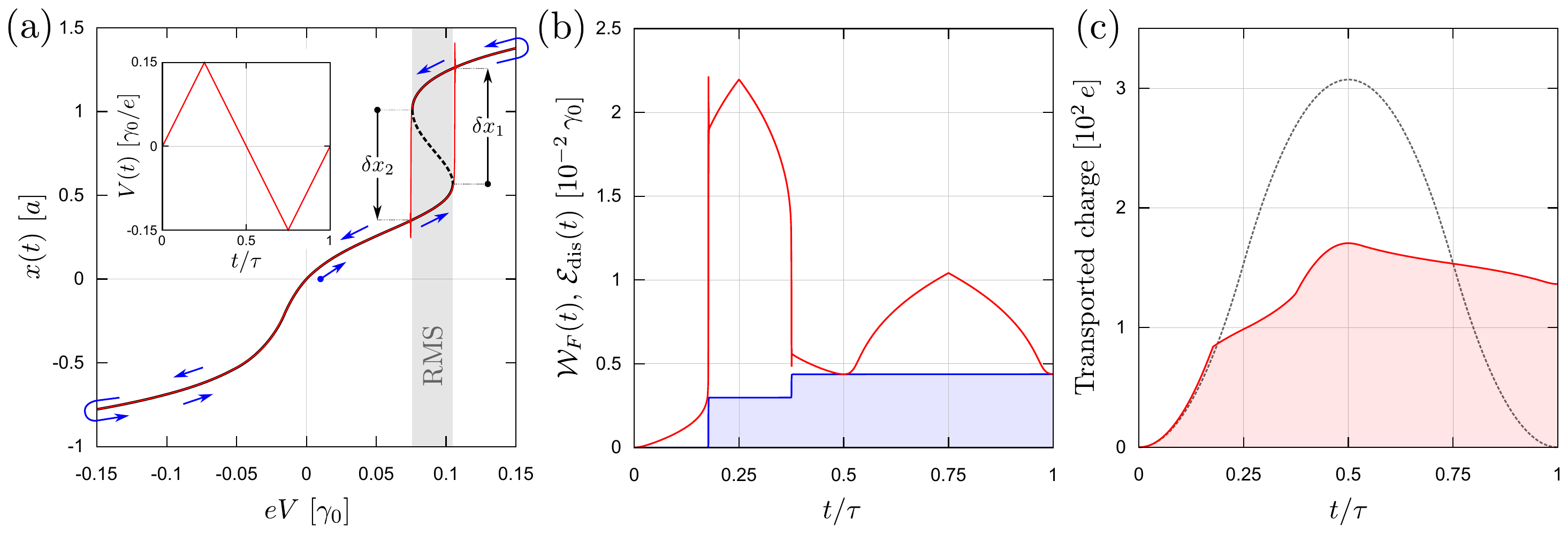}
\caption{Multiple stable solutions and hysteresis cycle. (a) QD's equilibrium position as a function of the bias voltage. The solution to Eq.~(\ref{eq:motion}) is shown in red, and the stationary solution is depicted in black. The shaded gray area denotes the bias region of multiple solutions, and the dashed line corresponds to the unstable positions. Black arrows indicate the jumps that the QD performs along its trajectory, which is denoted by the blue arrows. The inset shows the bias profile in a cycle from $t=0$ to $\tau$. (b) Work done by the CIF and dissipation energy calculated through Eqs.~(\ref{eq:wf}) and (\ref{eq:edis}) in red and blue, respectively, for intermediate times within the cycle of the bias voltage. (c) Transported charge as a function of time for $\lambda = 0.6$ (red) and $\lambda = 0$ (dashed black). The used parameters are $E_\text{dot}=-0.3\,\gamma_0$, $\gamma=0.2\,\gamma_0$, $\lambda = 0.6$ [except dashed black line in (c)], $k=0.02\,\gamma_0/a^2$, $b=7$, $a_0=4a$, and $\mu_0 = E_0 = 0$. For the dynamics we use $\tau = 10^6 \, \hbar/\gamma_0$, $M = 1000 \, \hbar^2/\gamma_0 a^2$, and $\zeta = 0.3$.} 
\label{fig:bias2}
\end{figure*}

As we already stated, under certain parameter conditions, it is possible to obtain more than a single stable solution for the QD position. This was shown, for example, in Fig.~\ref{fig:force_dyn}(b), where the equilibrium condition $F^\text{ne}(x_\text{st}) = k x_\text{st}$ is fulfilled for three different values of $x_\text{st}$. Indeed, the final position that the QD adopts not only depends on the electronic parameters (hopping amplitudes, bias voltage, etc.) but also on the initial condition and the mechanical parameters (e.g., oscillator mass, damping ratio).
Under this scenario, we can imagine a situation in which the QD is initially placed in its natural equilibrium position $x_0$. As we \textit{slowly} increase $V$, the QD will move according to its stable solution $x_\text{st}$ up to a region where more than a single position is allowed. This is shown in Fig.~\ref{fig:bias2}(a), where the region of multiple solutions (RMS) is marked by the shaded gray area. In this region, the QD will follow the solution that is closest to the one at the previous value of $V$, such that the trajectory $x_\text{st}(V)$ will resemble the one shown in solid red in the figure. 

If we further increase the bias such that it leaves this region, the QD will perform a jump $\delta x_1$ and continue on its way through the solution above the RMS. Now, by decreasing the bias from its maximum value $V = 0.15 \, \gamma_0/e$, the QD enters again into the RMS, but in this case, it will follow the upper stable solution. Finally, if we further decrease $V$, the QD leaves the RMS and performs a jump $\delta x_2$. This defines a hysteresis loop for the position of the QD and the CIF. Given that the CIFs in the forward and backward directions are different, a finite work $\mathcal{W}_F$ can be extracted from the cycle, i.e., energy is flowing from the electronic degrees of freedom towards the mechanical degree of freedom~\cite{bustos2019}. This can be calculated through:
\begin{equation}
\mathcal{W}_F (\tau)= \int_0^\tau \text{d}t \, F^\text{ne}[x(t)]\dot{x}(t),
\label{eq:wf}
\end{equation}
where $\tau$ is the period of the cycle of $V(t)$. Note that work can also be defined for intermediate times as $\mathcal{W}_F (t)$.

Since in the example analyzed in Fig.~\ref{fig:bias2} no external force (load) is included, i.e., $F_\text{ext}=0 \rightarrow \mathcal{W}_\text{ext}=0$, the energy gained by the mechanical degree of freedom is completely lost as dissipation, such that $\mathcal{W}_F(\tau) - \mathcal{E}_\text{dis} (\tau) = \mathcal{W}_\text{ext} (\tau) = 0$, where
\begin{equation}
\mathcal{E}_\text{dis} (\tau) = \int_0^\tau \text{d}t \, \nu\dot{x}^2(t).
\label{eq:edis}
\end{equation}
In Fig.~\ref{fig:bias2}(b) we show the above integrals for intermediate times $\mathcal{W}_F(t)$ (red) and $\mathcal{E}_\text{dis}(t)$ (blue) within a cycle of the bias voltage.
As discussed, a slow variation of the bias voltage (or $\delta T$) around a region of multiple stable solutions allows the QD system to operate as a motor. The key ingredient here is the jump $\delta x$ that the QD takes whenever it leaves the RMS. Since the QD can no longer adiabatically follow the time variation of the bias (there are no $x_\text{st}$ solutions infinitesimally close to the previous one), the system acquires a nonnegligible velocity, independently of the value of $\text{d}V/\text{d}t$.
There $x$ and $F^\text{ne}$ are no longer functions of $V$ solely, and
the integrals $\mathcal{E}_\text{dis}$ and $\mathcal{W}_F$ take a finite value even for an infinitesimal time interval (within the time scale of the voltage variation). This can also be seen in Fig.~\ref{fig:bias2}(b), where the two steps in $\mathcal{E}_\text{dis}$ correspond to the jumps $\delta x_1$ and $\delta x_2$. 

Since we use $\lambda \ne 0$ in this example, there is an imbalance in the charge current between the first and second half of the cycle, so a net transported charge is expected, according to Eq.~(\ref{eq:pump}). This is shown in Fig.~\ref{fig:bias2}(c), where $Q(\tau) \approx 140 \, e$ for $\tau = 10^6 \, \hbar/\gamma_0$. In this case, the amount of transported charge is much smaller than the one shown in Fig.~\ref{fig:bias1}(c) where there is no RMS. This difference mainly comes from the used coupling strength to the leads, which typically needs to be small ($\gamma \leq 0.2 \gamma_0$) to obtain a RMS for the specified range of bias voltage.

\subsection{Dynamical breaking of inversion symmetry}
\label{subsec:dynamical breaking}

\begin{figure*}[ht]
\includegraphics[width=\textwidth]{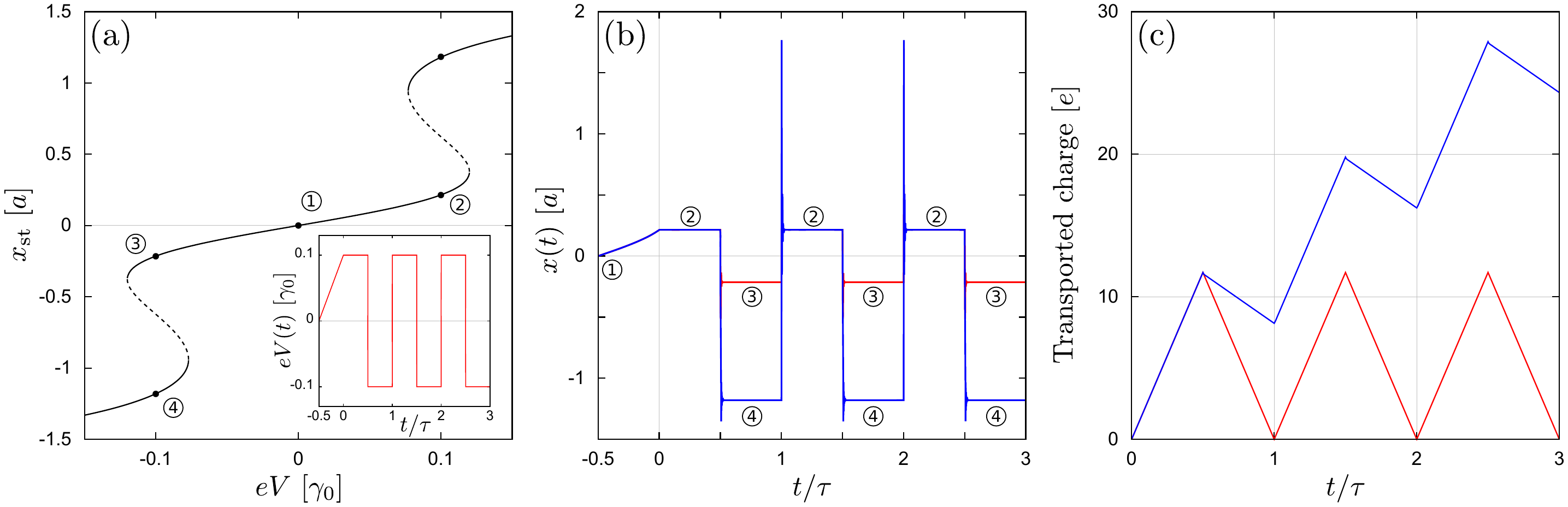}
\caption{Dynamical breaking of inversion symmetry. (a) QD's stationary position as a function of the bias voltage. The solid (dashed) line corresponds to stable (unstable) solutions. The numbered circles label the positions obtained in panel b for the time-dependent bias voltage shown in the inset. (b) Position of the QD as a function of time. For $-\tau/2 \leq t \leq 0$ the QD is slowly moved from its natural equilibrium position \Circled[]{1} to the lower solution \Circled[]{2}. Depending on the damping ratio $\zeta$, we obtain two dynamics: $\zeta = 0.2$ (red) and $\zeta = 0.16$ (blue). (c) Transported charge as a function of time for the two dynamics shown in (b). For this, we integrate the charge current according to Eq.~(\ref{eq:pump}) and take $t=0$ as the initial time. The used parameters are: $E_\text{dot}=-0.3\,\gamma_0$, $\gamma=0.1\,\gamma_0$, $\lambda = 0$, $k=0.02\,\gamma_0/a^2$, $b=6.5$, $a_0=4 a$, and $\mu_0 = E_0 = 0$. For the dynamics we use $\tau = 10^6 \, \hbar/\gamma_0$, and $M = 10^3 \, \hbar^2/\gamma_0 a^2$.} 
\label{fig:spont}
\end{figure*}

In this section, we discuss an interesting situation where an inversion-symmetric system, subjected to a symmetric modulation of $V$ (or $\delta T$), can exhibit rectification, in the sense that it can carry a net charge (or heat) per cycle.
This situation requires further exploitation of the nonlinear effect discussed above, to the point that multiple solutions appear even for $\lambda = 0$. Since the system preserves inversion symmetry, this effect is indeed related to a spontaneous breaking of symmetry due to the possibility of choosing between different steady-state positions through the QD dynamics. This is shown in Fig.~\ref{fig:spont}, where in panel (a) the stationary solution $x_\text{st}$ is displayed as a function of the bias voltage. In this case, there are two symmetric RMS for both positive ($V>0$) and negative ($V<0$) biases, each of them presenting two stable solutions.
It is clear that when $\lambda = 0$ a slow cyclic modulation of $V$ could not produce a net amount of charge current per cycle. However, this is not the case for a faster modulation of the voltage (fast with respect to the mechanical dynamics, though slow concerning the electronic degree of freedom).
In this scenario, depending on the initial condition, $\text{d}V/\text{d}t$, and the damping ratio $\zeta$, the system can stabilize in either of the two stationary solutions $x_\text{st}$ within the same RMS.
We therefore choose a modulation of $V$ such that it alternates its sign for a given value of $x_\text{st}$ within the RMS, i.e., $eV = \pm 0.1 \, \gamma_0$, see inset.
In addition, we initialize the position of the QD using a smooth ramp in the bias that goes from 0 to $0.1 \, \gamma_0$ for $-\tau/2 \leq t \leq 0$. This is shown in panel (b), where the QD is taken from its natural equilibrium position \Circled[]{1} to position \Circled[]{2} at time $t=0$.
After this time, we start the pulse sequence $eV/\gamma_0 = 0.1 \rightarrow -0.1 \rightarrow 0.1 \rightarrow \dots$, where the duration of the pulse is $\tau/2$. There we can see that for $\zeta = 0.2$ the QD adopts the symmetrical positions \Circled[]{2} and \Circled[]{3} (red line), while for a lower value $\zeta = 0.16$ (blue line) the negative position changes to \Circled[]{4} and, therefore, the movement is no longer symmetric. This is a consequence of the lower damping ratio, since the QD has enough kinetic energy to overcome the barrier separating the potential minima associated with \Circled[]{3} and \Circled[]{4}. 

Regarding the mean charge current per cycle, since $\lambda = 0$ we could naively expect that $I(-V) = - I(V)$. This would imply that the transported charge during the pulse with positive bias is compensated by that in the negative bias pulse when the two involved solutions are opposite in sign. Clearly, in this symmetric situation, the transported charge at the end of the cycle would be zero.
This is exactly what occurs for $\zeta = 0.2$ where the high friction compels the system to follow an ``adiabatic'' trajectory resulting in $x_\text{st}(V)=-x_\text{st}(-V)$. For $\zeta = 0.16$, however, the situation is different, and the solution chosen by the system within the same RMS depends on the QD velocity when entering that region. Interestingly, the solutions for the positive and negative voltage regions of the cycle are different, i.e., $x_\text{st}(V) \neq -x_\text{st}(-V)$.
This dynamical breaking of inversion symmetry results in a finite charge being transported, as can be seen in Fig.~\ref{fig:spont}(c).
Thus, the multiplicity of solutions offers the possibility of transporting a net charge between the contacts, even when the system preserves inversion symmetry and the bias modulation is symmetric and averages to zero over a period.
The direction of the net current is set by the initialization of the voltage modulation [see Fig.~\ref{fig:spont}(b) for $-\tau/2 \leq t 
 \leq 0$].~\footnote{In the inset of Fig.~\ref{fig:spont}(a) the voltage sequence starts at $t=0$ with a positive value. Indeed, the net current gets inverted if we change this initialization to a negative pulse.} In the next section we discuss the case of a noisy environment that could stochastically change the equilibrium positions from time to time. As we will see, one can always reset the system by a periodic initialization of the voltage modulation, to ensure the average direction of the mean current.

\textit{Comparison with similar systems.--} We should mention that there are other proposals involving electromechanical configurations similar to the one studied here. In particular, parametric instabilities have been studied in nanomechanical shuttles as a way of generating a dc current from ac voltages~\cite{ahn2006,penia2013}.
From the point of view of the models, the main difference with the cited works (besides the quantum treatment of the CIFs) is that we are applying square voltage pulses while they apply a sinusoidal time-periodic voltage in resonance with the frequency of the oscillator. However, this seemingly trivial difference leads to completely different mechanisms behind the charge rectification. 
In the mentioned works, as a result of the periodic perturbation, a parametric instability occurs, making the equilibrium position of the system unstable, thus generating a pair of bistable solutions~\cite{ahn2006}.
In our case, instead, the equilibrium positions of the QD are unaltered by the voltage, since we use a pulsed sequence where the modulus of $V$ remains fixed, and only its sign changes.
The difference also becomes evident by noticing that, for example, a parametric instability induces mechanical vibrations only when the frequency of the ac signal is close to the eigenfrequency of the mechanical subsystem~\cite{penia2013}. Alternatively, in cases such as the two shuttles used in Ref.~\cite{ahn2006}, parametric instabilities can occur when the oscillation frequencies are mode-locked to integer multiples of the applied voltage frequency. On the contrary, in our work we require a sudden change of the voltage which is instantaneous compared with the characteristic dynamics of the mechanical degree of freedom.
Moreover, in Ref.~\cite{penia2013} the direction of the current is determined by the phase shift between the ac gate voltage and the parametrically excited mechanical oscillations, while in our case the sign of the dc current depends on the initial conditions of the QD [set by the initialization phase of $V(t)$]. Another method to generate dc currents from a symmetric nanoelectromechanical shuttle is the one proposed in Ref.~\cite{qin2021} for a classical circuit. There, in addition to a sinusoidal ac bias voltage, periodic kicks are introduced and as a consequence, a rectified current arises due to the incommensurability between the shuttle motion and the driven bias, which also leads to chaotic motion. In contrast, in our case, the rectification is a result of a combination between the nonadiabatic response and the multiplicity of stable solutions.

Finally, it is important to emphasize that the effect described in this section exploits the nonadiabatic modulation of $V$ with respect to the mechanical degree of freedom. From the electron dynamics perspective, the voltage modulation is adiabatic in the sense that Eq.~(\ref{eq:g^<}) holds along the whole cycle of $V(t)$. Therefore, the phenomenon is completely different from quantum pumping~\cite{brouwer1998,bode2011}, which necessarily involves beyond-Born-Oppenheimer corrections to the current operator.

\begin{figure*}[ht]
\includegraphics[width=0.9\textwidth]{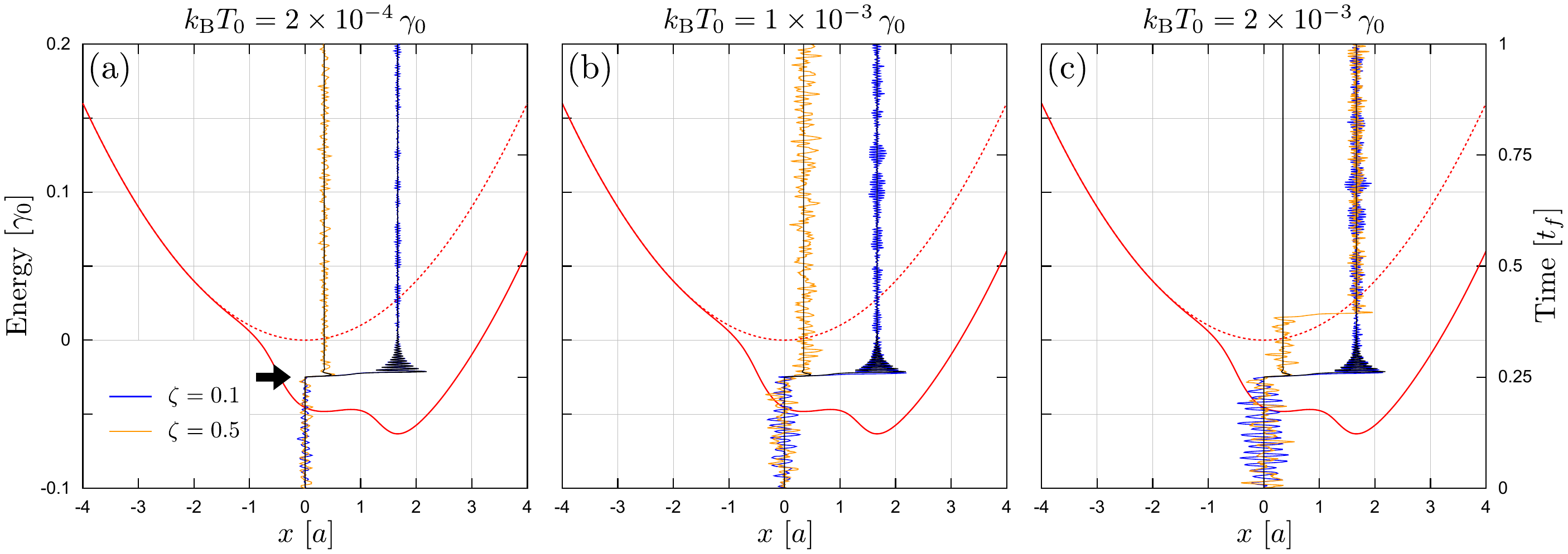}
\caption{Dynamics of the QD including the force fluctuation $\xi$ for $k_\text{B}T_0 = 2 \times 10^{-4} \, \gamma_0$ (a), $1 \times 10^{-3} \, \gamma_0$ (b), and $2 \times 10^{-3} \, \gamma_0$ (c). We use the damping ratios $\zeta = 0.1$ (blue) and $0.5$ (orange), and the black curves depict the case without fluctuations ($k_\text{B} T_0 = 0$). The black arrow in (a) marks the moment $t_f/4$ in which the bias voltage is turned on. The effective and harmonic energy potentials are shown in solid and dashed red, respectively. In all panels, the left axis shows the energy scale for the potentials, and the right axis corresponds to the evolution time, with $t_f = 10^5 \, \hbar/\gamma_0$. The other parameters coincide with those used in Fig.~\ref{fig:force_dyn}(b).} 
\label{fig:dyn_xi}
\end{figure*}

\subsection{Multistability and force fluctuations}\label{sec:fluctuation}

Here we discuss the role of the fluctuation term of the forces in the proposed electromechanical system. In particular, by analyzing the dynamics of the QD system under the influence of non-zero force fluctuations, we investigate the dynamical breaking of the inversion symmetry discussed in Sec.~\ref{subsec:dynamical breaking}.

To begin with, we will consider that the force fluctuation $\xi$ is dominated by its equilibrium contribution, given that the nonequilibrium component is related to a higher-order term in the force expansion in both the QD velocity and the voltage/temperature bias~\cite{bode2011,bustos2019,deghi2021}. If we assume that the autocorrelation function of the stochastic mechanical force is local in time, $\braket{\xi(t)\xi(t')} = D \delta(t-t')$, we can relate, in equilibrium, the zero frequency noise $D$ with the mechanical friction coefficient $\nu$ through the fluctuation-dissipation theorem:
\begin{equation}
D = 2 k_\text{B}T_0 \nu.
\end{equation}
In such a case, fluctuations should be small compared to the remaining forces in the low-temperature limit. However, as we will see below, how small the temperature should be, depends on the parameter regime we are considering. To inspect this, we include the fluctuation force $\xi$ in Eq.~(\ref{eq:langevin}) through:
\begin{equation}
\xi(t) = g_t \sqrt{\frac{D}{\Delta t}},
\end{equation}  
where $g_t$ is a random value extracted from a Gaussian distribution and $\Delta t$ represents the time step used in the integration algorithm~\cite{calvo2017}.

In Fig.~\ref{fig:dyn_xi} we recalculate the dynamics of the QD related to the situation shown in Fig.~\ref{fig:force_dyn}(b) for positive bias, where the system presents bistability. For comparison, we also show the effective potential energy surface, defined as:
\begin{equation}
\mathcal{U}_\text{eff}(x) = \mathcal{U}_\text{eq}(x) + \mathcal{U}_\text{ne}(x)  = \frac{kx^2}{2} - \int_{-a_0}^{x} F^\text{ne}(x') \text{d}x',
\end{equation}
which in this case develops a double-well shape. To initialize the system, we set $V=0$ for $0 \leq t \leq t_f/4$, where $t_f = 10^5 \, \hbar/\gamma_0$ is the final evolution time. During this interval, the QD position is confined in a region $\Delta x$ around $x_0 = 0$, given by $k_\text{B}T_0$ and the force constant $k$. For $t>t_f/4$ the bias voltage is turned on, and the QD will move into one of the two wells of $\mathcal{U}_\text{eff}$. In the examples shown in Figs.~\ref{fig:dyn_xi}(a) and (b) we can see that, depending on the damping ratio $\zeta$, the QD falls into one of the two wells and remains there during the entire time evolution. In particular, for $\zeta = 0.5$ the QD falls into the local minimum associated with $x_{\text{st},1}$ of Fig.~\ref{fig:force_dyn}(b). However, when the thermal energy becomes comparable to the potential barrier of this well, i.e., the energy difference $\Delta \epsilon$ between the local maximum and the local minimum at $x_{\text{st},1}$ (in this example $\Delta \epsilon \approx 1.3 \times 10^{-3} \, \gamma_0$), the dynamics is no longer stable. Now, there is a non negligible probability for the QD to overpass the barrier and fall into the global minimum at $x_{\text{st},2}$, as happens in Fig.~\ref{fig:dyn_xi}(c), where $k_\text{B}T_0 \approx 1.5 \Delta \epsilon$. Therefore, if we use $\gamma_0 = 3.88$ eV as in Appendix~\ref{app:unit}, the temperature $T_0$ below which the QD remains in $x_{\text{st},1}$ should be considerably smaller than $58$ K. The condition $k_\text{B}T_0 \ll \Delta \epsilon$ therefore ensures bistability under force fluctuations, at least for a finite evolution time $t_f$.

\begin{figure*}[ht]
\includegraphics[width=0.9\textwidth]{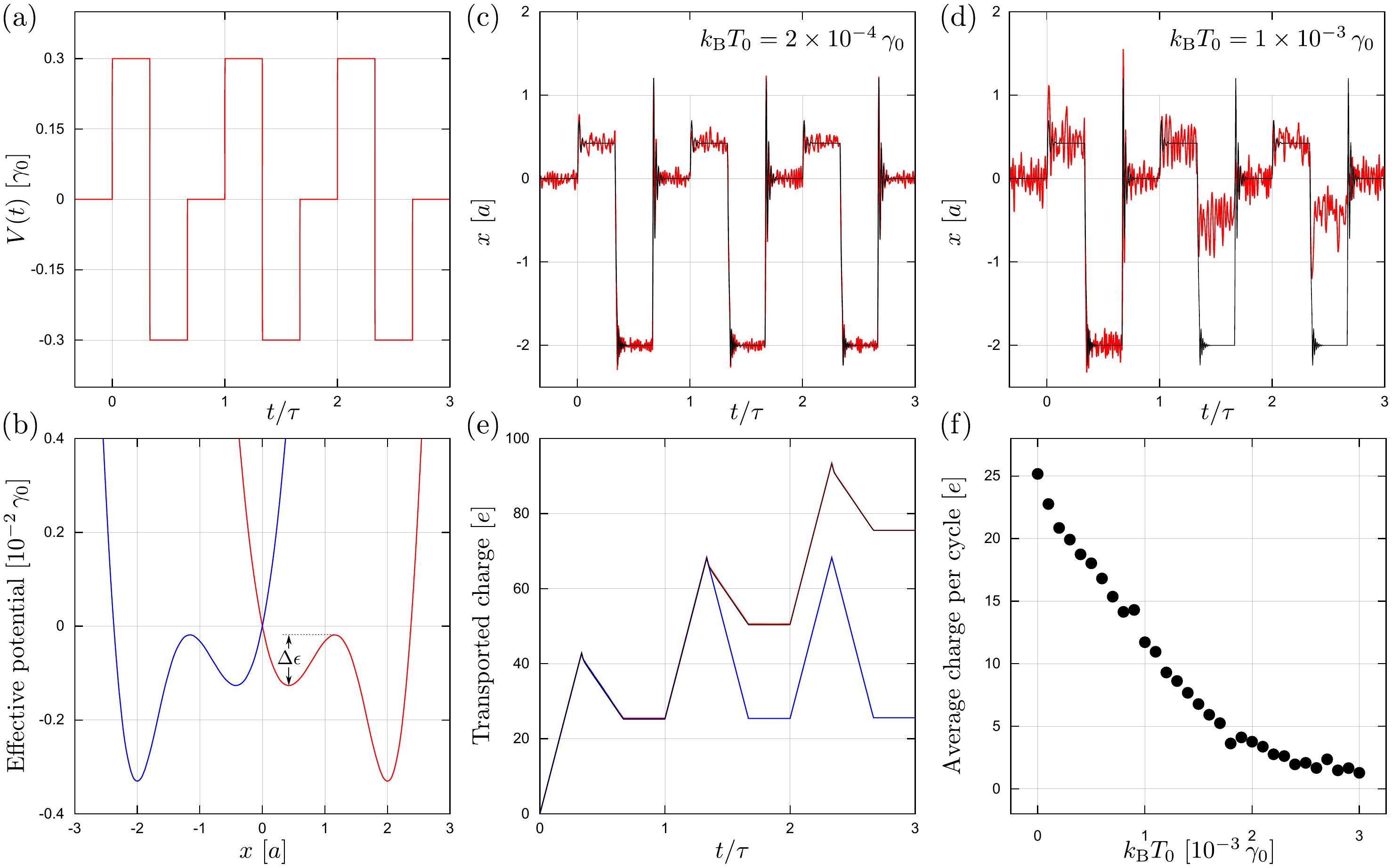}
\caption{Dynamical breaking of inversion symmetry including force fluctuations. (a) Bias voltage as a function of time, including the initialization from $t=-\tau/3$ to $t=0$. The figure displays three cycles with $\tau = 2/3 \times 10^5 \, \hbar/\gamma_0$. (b) Effective potential energy $\mathcal{U}_\text{eff}(x)$ for $V = 0.3 \, \gamma_0$ (red) and $V = -0.3 \, \gamma_0$ (blue). The energy barrier is $\Delta \epsilon \approx 10^{-3} \, \gamma_0$. (c) and (d) QD position as a function of time for $k_\text{B}T_0 = 2 \times 10^{-4} \, \gamma_0$ and $10^{-3} \, \gamma_0$, respectively, in solid red. (e) Transported charge as a function of time for the examples shown in panel c (red) and d (blue). In (c-e) the solid black curves show the zero-temperature limit. (f) Transported charge per cycle as a function of the thermal energy, averaged over 1500 cycles. Parameters: $E_\text{dot}=-0.3\,\gamma_0$, $\gamma=0.11\,\gamma_0$, $\lambda = 0$, $k=0.02\,\gamma_0/a^2$, $b=6.5$, $a_0=7.5 a$, $\mu_0 = E_0 = 0$, $M = 10^3 \, \hbar^2/\gamma_0 a^2$, and $\zeta = 0.16$.} 
\label{fig:spont_xi}
\end{figure*}

We now address the inclusion of force fluctuations in the dynamical breaking of inversion symmetry discussed in Sec.~\ref{subsec:dynamical breaking}. As the required bistability can stochastically be broken through the thermal energy, we would like to establish some temperature threshold to ensure that this nonlinear effect survives at least on average. This is  explored in Fig.~\ref{fig:spont_xi}, where we employ a similar sequence for the voltage pulses as that shown in Fig.~\ref{fig:spont}. Here, in addition to the square pulses with finite bias, we also include a third pulse with $V=0$ to ensure the QD initiates, before the next cycle, in a well-defined average position (i.e., $\bar{x}=x_0$). Importantly, with this modification, the bias voltage still presents a zero-mean value over a cycle. Another difference is that, in the present case, we do not use an initial ramp for the bias, as this should not be taken into account when taking averages over trajectories. In Fig.~\ref{fig:spont_xi}(c) we can see the typical dynamics for low temperatures, where the QD moves according to the trajectory dictated by $T_0 = 0$ (solid black), displaying small deviations that do not affect the sequence of stable positions [associated with the minima shown in panel (b)]. In particular, the amount of transported charge remains unaffected in this case, as can be noticed in panel (e). For a larger temperature, as shown in panel (d), the QD no longer performs a well-defined sequence, in the sense that between pulses it may not fall into the correct potential well. By ``correct'' we mean the potential well the QD falls in the zero temperature limit. Importantly, this effect is not the same as the one observed in Fig.~\ref{fig:dyn_xi}(c), since the QD remains around the same potential well while the voltage is kept constant. Here, the fluctuation in the force leads to deviations in the QD velocity, preventing it from staying in the required range $\Delta \dot{x}$ to ensure that the correct potential well is reached. When this occurs, the transported charge in this particular cycle is canceled out, thus reducing the total transferred charge during the entire sequence, see blue curve in (e). The other effect, related to the one what was observed in Fig.~\ref{fig:dyn_xi}(c), is that even within the pulse, the QD may perform a jump to the other potential well, and in this cycle, the transferred charge can be inverted. In other words, these two effects reduce the total transferred charge since (1) bistability is no longer ensured and (2) the QD is not able to dissipate the correct amount of kinetic energy. This is shown in Fig.~\ref{fig:spont_xi}(f) where we calculate the average transferred charge per cycle over 1500 cycles as a function of the thermal energy, where a sustained decay is observed. 

It is important to highlight that, although the obtained temperature range for the nonlinear effects to become evident is relatively small, this corresponds to a particular example with a given set of parameters in a minimal model. Nevertheless, this is not necessarily a restriction since, e.g., larger potential barriers and/or more confining wells can be obtained, such that the nonlinear effects would become more robust to larger temperatures.

\section{Conclusions}
\label{sec:conclusions}

By using a generic minimal model, we have explored the potential of CIFs to introduce nonlinearities in the transport properties of a wide range of composite systems, enabling for the rectification of heat and charge currents, and the emergence of hysteretic behavior in molecular and nanoelectromechanical devices.
We have proved that even simple systems, like the one studied here, can show important rectification characteristics with rectification factors close to the unity. Under the appropriate parametrization, we have also found that the system can adopt multiple stable positions allowing for work extraction under an adiabatic (time-periodic) modulation of the bias voltage (or temperature gradient).

Additionally, we have shown how the multiplicity of steady-state solutions can be used to pump charge (or heat) in symmetric systems modulated by symmetric time-dependent voltage biases (or temperature gradients). We have rationalized this, which contradicts the common intuition on pumping within condensed matter physics, as the result of a symmetry breaking allowed by the system's dynamics when it cannot adiabatically follow a sudden bias variation. 

Finally, we have investigated the role of the force fluctuations in the dynamics of the electromechanical system. We observe that nonlinear effects associated with multiple stable solutions endure, on average, as long as the thermal energy remains below the energy barriers separating the minima of the effective potential. Although the temperature range required for this phenomenon is relatively narrow, a comprehensive exploration of the parameters is yet to be implemented. 

Our results can lead to the development of new nanoscale (or molecular) rectifiers, as well as novel forms of pumps and motors, which constitute building blocks for nanoscale heat management, energy harvesting, and nanometric actuators. In this regard, we believe that the present work paves the way for exploring the discussed nonlinear phenomena in more realistic scenarios, for example, by using concrete molecular models for the local system, or including finite temperature effects on the system's dynamics, beyond the linear response regime. 
Other interesting research directions to extend our work include the use of superconducting leads to study supercurrent rectification in similar models~\cite{zazunov2010}, as well as assessing the role of quantum effects on the mechanical coordinate~\cite{erpenbeck2018} and its impact on reciprocity breaking and multistability.

\section{Acknowledgements}
The authors are thankful for the financial support from Consejo Nacional de Investigaciones Científicas y Técnicas (CONICET, PIP-2022-59241); Secretaría de Ciencia y Tecnología de la Universidad Nacional de Córdoba (SECyT-UNC, Proyecto Formar 2020); and Agencia Nacional de Promoción Científica y Tecnológica (ANPCyT, PICT-2018-03587). All authors are members of CONICET and thank Luis E. F. Foa Torres for fruitful discussions and hospitality during the 2023 ICTP conference at Universidad de Chile, Santigo de Chile.

\appendix

\section{Units of the physical quantities}
\label{app:unit}

In order to deduce the typical scales involved in our quantities $R$, we write $R = R^* \tilde{R}$, where $R^*$ corresponds to the unit of $R$ and $\tilde{R}$ represents its dimensionless value. For the energy, we will use the hopping $\gamma_0$ as the unit of reference, i.e., $\epsilon^* = \gamma_0$, such that we can take the time unit from $\hbar/t^* = \gamma_0$ and define $t^* = \hbar/\gamma_0$. For the position of the QD, we will use the lattice constant $a$ as the unit of length. Although this would seem restricted to leads composed by one-dimensional homogeneous chains, we can generalize it from the definition of the Fermi velocity, such that $a \simeq v_\text{F} t^* = \hbar v_\text{F}/\gamma_0$. These are the main scales of the system and through them we can derive the remaining ones. For example, for the CIF we use its definition as $F^\text{ne} = -\braket{\partial_x \hat{H}_\text{el}}$, such that $F^* = \gamma_0/a$. By using these scales, we can rewrite Eq.~(\ref{eq:motion}) in terms of the dimensionless quantities as follows:    
\begin{equation}
\tilde{M} \frac{\text{d}^2 \tilde{x}}{\text{d} \tilde{t}^2} + \tilde{k}\tilde{x} + \tilde{\nu} \frac{\text{d} \tilde{x}}{\text{d} \tilde{t}^2} = \tilde{F}^\text{ne},
\end{equation}
where we obtain $M^* = \hbar^2/(\gamma_0 a^2)$, $k^* = \gamma_0/a^2$, and $\nu^* = \hbar/a^2$. Notice, for example, that the oscillator energy is also referred to $\gamma_0$, i.e., $\hbar\omega^* = \hbar \sqrt{k^*/M^*} = \gamma_0$, and the damping ratio $\zeta$ remains dimensionless. For example, by taking $\gamma_0 = 3.88$ eV and $a = 2.5$ \AA~\cite{dundas2009}, we obtain an elastic constant unit $k^* \approx 10$ N/m such that for typical van der Waals forces between fullerene molecules~\cite{cox2007} where $k = 0.23$ N/m, its dimensionless value would be $\tilde{k} = k/k^* \approx 0.02$. Regarding the mass unit, for these parameters we obtain $M^* \approx 0.3 \, m_\text{e}$, with $m_\text{e}$ being the electron mass.
From the above considerations, it becomes evident that $\omega=\sqrt{\tilde{k}/\tilde{M}}\gamma_0/\hbar$, leading to an effective frequency of the mechanical motion at least 22 times smaller than $\gamma/\hbar$ for $\tilde{k}=0.02$, $\tilde{M}=1000$ and $\gamma=0.1 \, \gamma_0$. Moreover, in some cases, the frequency can be up to 110 times smaller when $\gamma=0.5 \, \gamma_0$. Since $\omega \ll \gamma/\hbar$, this enables a time-scale separation between the fast electronic relaxation and the slow vibrational motion, allowing to assume an adiabatic condition. In any case, as time progresses, the velocity of the QD approaches zero and hence the adiabatic condition will inevitably be satisfied. In this sense, the QD stable positions can be found regardless of the details in the initial dynamics. 

\section{Analytic expression for the current-induced force}
\label{app:formula}

In the linear response regime, characterized by a low bias voltage and low temperature, we have the following equation for the force induced by a bias voltage, see Eq.~(\ref{eq:cif_V})
\begin{equation}
F^\text{ne}_{\delta\mu} = \sum_\alpha \frac{\delta\mu_\alpha}{\pi}\text{Tr}[\bm{\Lambda}\bm{G}^r \bm{\Gamma}_\alpha \bm{G}^a]_{\mu_0},
\end{equation} 
where the $\alpha$ index accounts for left and right leads and we assume a symmetric shift of the electrochemical potentials, such that $\delta \mu_\text{L} = - \delta \mu_\text{R} = eV/2$. The matrix representation of the force operator in the site basis is:
\begin{equation}
\bm{\Lambda} = \frac{b}{a_0} \begin{pmatrix}
0 & -\gamma_\text{L} & 0\\
- \gamma_\text{L} & 0 & \gamma_\text{R}\\
0 & \gamma_\text{R} & 0
\end{pmatrix},
\end{equation}
in addition, $\bm{\Gamma}_\alpha = (\bm{\Sigma}_\alpha^a-\bm{\Sigma}_\alpha^r)/2i$ can be obtained from the self-energy of reservoir $\alpha$, and it has the matrix representation: $\bm{\Gamma}_\text{L} = \Gamma \, \text{diag}(1,0,0)$, and $\bm{\Gamma}_\text{R} = \Gamma \, \text{diag}(0,0,1)$, where $\Gamma$ represents the escape rate from the system to the lead. Evaluated at the Fermi energy $\mu_0 = E_0 = 0$, this function takes the value $\Gamma = \gamma_0$. We therefore obtain:
\begin{align}
F^\text{ne}_{\delta\mu} = &\frac{\gamma_0 eV}{\pi} \left(\frac{b}{a_0}\right) \left[\gamma_\text{L} \text{Re}(G_{13}G_{23}^*-G_{11}G_{21}^*) \right. \nonumber \\
& \left. +\gamma_\text{R} \text{Re}(G_{21}G_{31}^*-G_{33}G_{23}^*)\right],
\end{align}
where we omit the $r$ supraindex in the $G$'s as all matrix elements are referred to the retarded Green's function, since we use $\bm{G}^a = [\bm{G}^r]^\dag$. For the considered model of Eq.~(\ref{eq:Heff}), the matrix elements of the retarded Green's function $\bm{G}^r$ are calculated in Ref.~\cite{deghi2021} (see Appendix D), where we can use the following relations:
\begin{align}
\text{Re}(G_{11} G_{21}^*) &= \frac{\gamma_\text{L} E_\text{dot}}{(E_\text{dot}\gamma_0)^2+(\gamma_\text{L}^2+\gamma_\text{R}^2)^2}, \\
\text{Re}(G_{33} G_{23}^*) &= \frac{\gamma_\text{R} E_\text{dot}}{(E_\text{dot}\gamma_0)^2+(\gamma_\text{L}^2+\gamma_\text{R}^2)^2},
\end{align}
and $\text{Re}(G_{13}G_{23}^*) = \text{Re}(G_{21}G_{31}^*) = 0$. Replacing them in the above equation yields the following analytic expression for the CIF:
\begin{equation}
F^\text{ne}_{\delta\mu}(x) = -\frac{eV}{\pi}\left(\frac{b}{a_0}\right)\frac{(E_\text{dot}\gamma_0)(\gamma_\text{L}^2+\gamma_\text{R}^2)}
{(E_\text{dot}\gamma_0)^2+(\gamma_\text{L}^2+\gamma_\text{R}^2)^2}.
\end{equation}
By considering the explicit form of the hopping amplitudes to the leads given in Eq.~(\ref{eq:hop}), we obtain that the $x$-dependence on $F^\text{ne}$ is of the form:
\begin{equation}
\gamma_\text{L}^2+\gamma_\text{R}^2 = 2\gamma^2 \left[ (1+\lambda^2) \cosh y - 2\lambda \sinh y \right],
\end{equation}
where $y=2bx/a_0$ is the (dimensionless) length scale that is relevant for the CIF.

\bibliography{cite}

\end{document}